\documentclass[aps,prl,twocolumn,groupedaddress,amsmath,amssymb]{revtex4}
\usepackage{graphicx}
\usepackage{dcolumn}
\usepackage{bm}
\usepackage{hyperref}

\newcommand{\be}{\begin{equation}}
\newcommand{\ee}{\end{equation}}
\newcommand{\bea}{\begin{eqnarray}}
\newcommand{\eea}{\end{eqnarray}}
\begin{document}

\title{Gravity Probe B: Final Results of a Space 
Experiment to Test General Relativity}

\author{C.~W.~F.~Everitt$^{1*}$,
D.~B.~DeBra$^1$, B.~W.~Parkinson$^1$, J.~P.~Turneaure$^1$, \\
J.~W.~Conklin$^1$,  M.~I.~Heifetz$^1$, G.~M.~Keiser$^1$,  A.~S.~Silbergleit$^1$,\\
T.~Holmes$^1$, J.~Kolodziejczak$^2$, M.~Al-Meshari$^3$, J.~C.~Mester$^1$, B.~Muhlfelder$^1$, V.~Solomonik$^1$, K.~Stahl$^1$, P.~Worden$^1$,\\
W.~Bencze$^1$, S.~Buchman$^1$, B.~Clarke$^1$, A.~Al-Jadaan$^3$, H.~Al-Jibreen$^3$, J.~Li$^1$, J.~A.~Lipa$^1$, J.~M.~Lockhart$^1$,\\
B.~Al-Suwaidan$^3$, M.~Taber$^1$, S.~Wang$^1$}
\affiliation{$^1$HEPL, Stanford University, Stanford, CA 94305-4085, USA; $^*${\bf e-mail:} {francis1@stanford.edu}\\
  $^2$ George C. Marshall Space Flight Center, Huntsville, AL 35808, USA\\ 
  $^3$ King Abdulaziz City for Science and Technology, Riyadh, Saudi Arabia}
\date{\today}

\begin{abstract}
Gravity Probe B, launched 20 April 2004, is a space experiment testing
two fundamental predictions of Einstein's theory of General Relativity (GR),
the geodetic and frame-dragging effects,
by means of cryogenic gyroscopes in Earth orbit.
Data collection started 28 August 2004 and ended 14 August 2005.
Analysis of the data from all four gyroscopes results in a geodetic drift rate
of $-6,601.8\pm 18.3$ mas/yr and a
frame-dragging drift rate of $-37.2 \pm 7.2$ mas/yr,
to be compared with the GR predictions of
$-6,606.1$ mas/yr and $-39.2$ mas/yr, respectively (`mas' is milliarc-second; $1\,$mas\,$=4.848\times10^{-9}$ rad).
\end{abstract}

\pacs{}

\maketitle

\section{Introduction}

In 1960, L.~I.~Schiff  \cite{schiff1960} showed that an ideal gyroscope
in orbit around the Earth would undergo two relativistic precessions
with respect to a distant inertial frame:
1)~a geodetic drift in the orbit plane due to motion through the space-time curved by the Earth's mass;
2)~a frame-dragging due to the Earth's rotation.
The geodetic term matches the curvature precession of the
Earth-Moon system around the Sun given by
W.~de~Sitter in 1916~\cite{deSitter1916}.
The Schiff frame-dragging is related to
the dragging of the orbit plane of a satellite around a rotating planet
computed by J.~Lense and H.~Thirring in 1918~\cite{lense1918pz}. 
Frame dragging has important implications for astrophysics; it has been invoked as a mechanism to drive relativistic jets emanating from galactic nuclei~\cite{thorne1988}. 
\begin{figure}[b]
  \includegraphics[width = 8 cm]{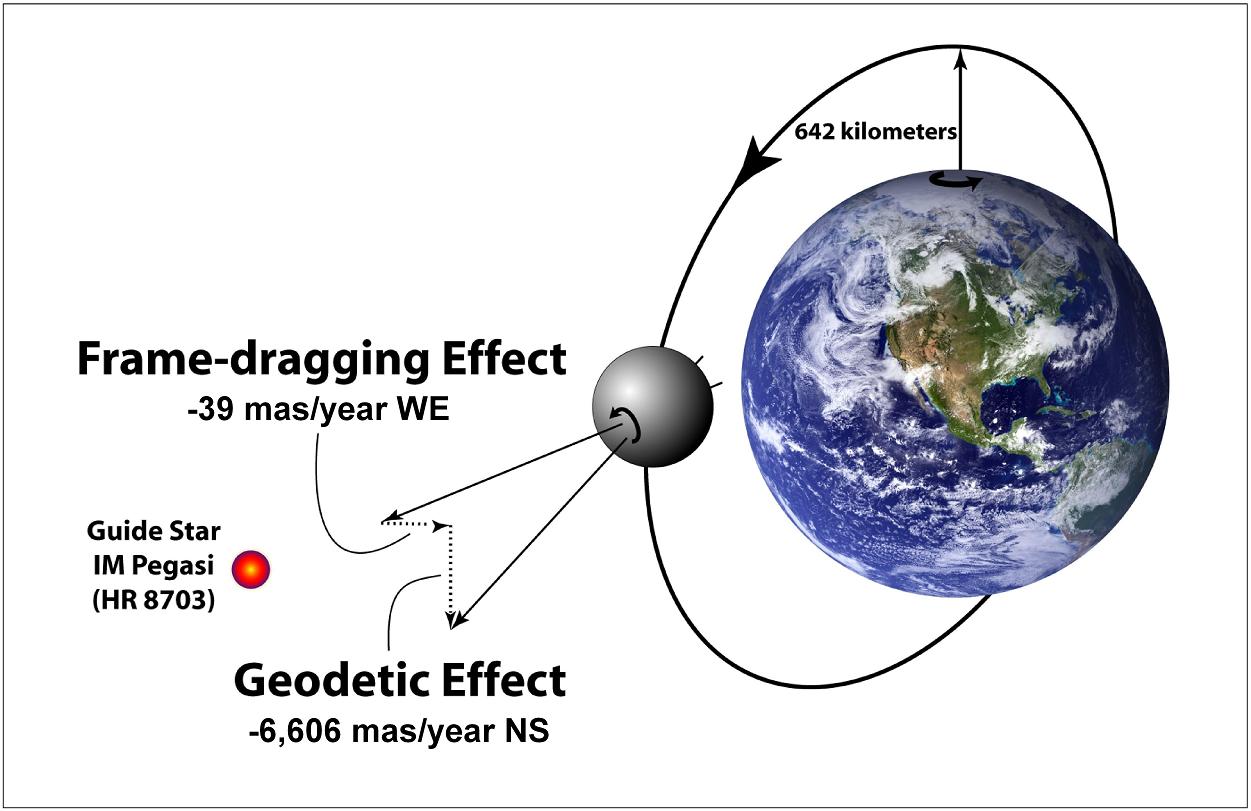}
  \caption{Predicted drift rates of GP-B gyroscopes.\\  See~\cite{footnote1} for definitions of WE and NS inertial directions.
         \label{figOverview}}
\end{figure}

The measurement requires one or more gyroscopes referenced to a remote star
by an onboard telescope.
In the 642 km polar orbit of Gravity Probe B,
the two effects are at right angles, as in Fig. \ref{figOverview}. 
The predicted geodetic drift rate is $-6,606.1$ mas/yr;
the frame-dragging drift rate with the chosen star IM Pegasi
is $-39.2$ mas/yr.

GP-B was conceived as a controlled physics experiment
having mas/yr stability ($10^6$ times better than the best modeled navigation
gyroscopes) with numerous built-in checks and methods of treating systematics.
Three principles guided the design:
1) make Newtonian gyro drifts $\ll$ 
the predicted GR effects;
2) add sensors so that modeling hinges on physical
understanding as against the \emph{ad hoc}
observational modeling used in navigation gyroscopes;
3) exploit natural effects such as stellar aberration in
calibrating the instrument. Meeting the
many  mechanical, optical, and electrical requirements rested
on a conjunction of two technologies, space and cryogenics.

Operation in space separates the two effects, increases the geodetic effect $12.4$ times as compared to a gyroscope at the equator,
 eliminates `seeing' in the measurement to the guide star, and vastly reduces torques from suspending the gyroscope against
$1\,$g gravity. Cryogenics brings new levels of magnetic shielding, thermal isolation,
ultra-high vacuum operation, and a uniquely effective gyro readout based
on the London moment in a spinning superconductor.
The two together give the instrument ultimate mechanical stability:
in zero $g$, there is no sag; at zero K, there is no thermal 
distortion.

Essential to GP-B as a controlled physics experiment was the calibration phase, a 46-day period 
following the main science phase designed to set limits on a range of potential disturbing effects 
and quantify any that might prove larger than expected. 

\section{Experiment Description}

The heart of the instrument was a $0.92\,$m long fused silica structure
containing four gyroscopes and a star-tracking telescope mounted in a
$2440\,\ell$ superfluid helium dewar operating at $1.8\,$K.
Each gyroscope comprised a $38\,$mm diameter niobium-coated fused quartz
sphere suspended electrically, spun up by helium gas,
and read out magnetically. The four were set in line,
two spinning clockwise and two counterclockwise,
with axes initially aligned to the boresight of the telescope.
The space vehicle rolled with 77.5 s period about the line of sight to IM Pegasi,
which was, however, occulted by the Earth for almost half each orbit
(Fig. \ref{figOverview}).
During occulted periods, pointing was referenced to star trackers and rate
gyros on the outside of the spacecraft.
Drag compensation, originated by G.~E.~Pugh in an independent proposal for an orbiting gyroscope experiment~\cite{pugh1959}, two months prior to Schiff's paper, was by a control system referred to one of the gyroscopes as a proof mass.
Attitude, translational, and roll control authority was provided
by the helium boil--off gas from the dewar vented through
proportional thrusters~\cite{debra1988}.

Vital was a gyro readout that did not disturb the spin orientation.
Superconductivity supplied three essentials.
The spinning rotor generated a London moment equivalent at 80 Hz to a uniform $5\times10^{-5}$G field aligned with the spin axis. 
A SQUID magnetometer coupled to a superconducting loop on the gyro housing provided a readout capable
of detecting a $1\,$mas change of spin direction in $10\,$hr. Finally, a combination of high permeability and ultra-low field superconducting shields around the instrument with local shields for each gyro achieved: 1) 240 dB isolation from external magnetic disturbances; 2) a limit on trapped fields in the rotors $\sim\,$1\% of the London moment; and 3) virtual elimination of magnetic torques.

The gyro readout scale factor $C_g(t)$ was calibrated on orbit against the aberration of starlight. During  each half--orbit when IM Pegasi was visible, the telescope pointed at its apparent position, and the $20.49586\,$arc-s annual and $5.18560\,$arc-s orbital aberrations, derived respectively from the JPL Earth ephemeris and GPS orbit data, appeared in the gyro readout.
\begin{table}[t]
  \caption{\label{tabReq} The seven near zeroes}
  \begin{ruledtabular}
    \begin{tabular}{l l l}
      Property & Requirement & Achieved \\
      \hline \\[-2ex]
      \textbf{Rotor Properties} & & \\
      Mass unbalance, $\delta r /r$ & $6\times10^{-7}$ & $2-5\times10^{-7}$ \\
      Asphericity (nm) & $\quad55$ & $\quad<33$ \\
      Patch dipole (C-m) & $< 10^{-15}$& $\quad$\emph{see text} \\
      \textbf{Environment} & & \\
      Cross-track acceleration (g) & $\quad 10^{-11}$ & $\qquad10^{-11}$ \\
      Gas pressure (torr) & $\quad 10^{-11}$ & $\quad< 10^{-14}$ \\
      Rotor trapped field ($\mu$G) & $\quad\; 9$ & $\quad0.2 - 3$ \\[1ex]
      \textbf{Mixed} & & \\
      Rotor electric charge (electrons) & $\quad 10^8$ & $\quad <10^8$
    \end{tabular}
  \end{ruledtabular}
\end{table}

An elementary calculation captures the stringencies of the experiment.
Consider a spherical, not quite homogeneous rotor under transverse acceleration $f$.
Let $\delta r$ separate the centers of mass and support, $v_s\sim9$ m/s be the rotor's peripheral velocity,
$\Omega_0$ the maximum allowed drift rate $0.1\,\mbox{mas/yr}$ ($1.5\times10^{-17}$rad/s).
Then, we have $\delta r/r<2v_s\Omega_0/5f$, and mass unbalance requirements $\delta r/r$:
$6\times10^{-18}$ on Earth;
$6\times10^{-10}$ in a~typical 642~km altitude satellite;
and $6\times10^{-7}$ for GP-B in $10^{-11}$ g drag-free mode.
Without drag compensation, the experiment would have been impossible. Likewise for an aspherical homogeneous rotor, the electrical support torques, while greatly reduced on orbit, only reached the desired level through a further symmetrizing factor, spacecraft roll. The net allowed asphericity was $55\,$nm.
\begin{figure}[b]
 \includegraphics[width = 8 cm]{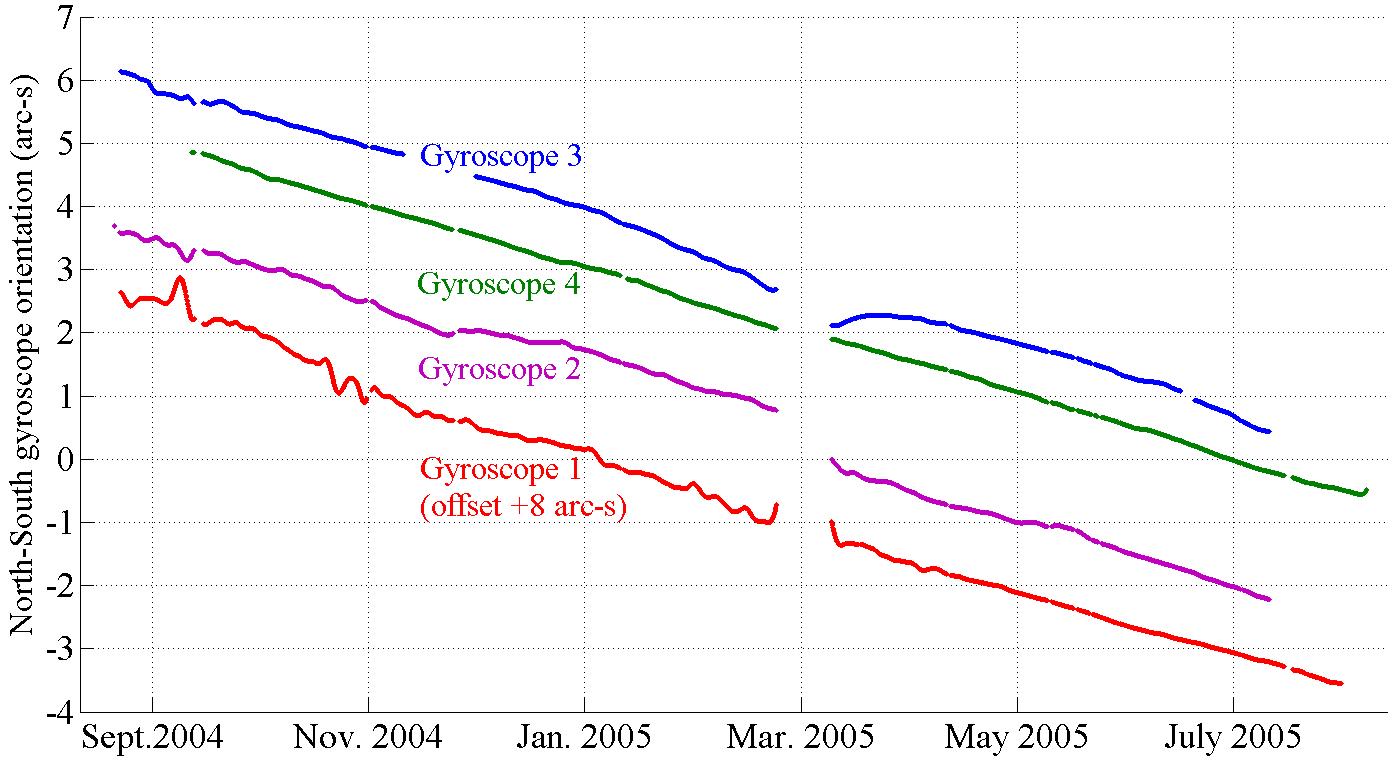}
  \caption{\label{figNS} North-South gyro orientation 
    histories with no modeling of torque or scale factor}
\end{figure}

Aiming for a  $\sim 0.5\,$mas/yr mission, we created an error tree setting limits on 133 disturbing terms, and verified in advance that all were negligible. Central were seven `near zeroes' (Table \ref{tabReq}). Four noted already are inhomogeneity and asphericity for the rotor, residual acceleration and magnetic field for the environment. The others are gyro electric charge, gas pressure, and patch effect. Six met requirements; the complication discovered during the calibration phase was the patch effect. Even so, the idea of seeing relativity in the `raw' data was preserved. Fig. \ref{figNS} shows the NS drifts of the four gyroscopes with no torque modeling. The geodetic drift is visible in all four.

The term `patch effect' \cite{darling1989, speake1996cqg} refers to contact potential differences between crystalline or contamination regions on a metal surface. Prelaunch studies focused on eddy current losses and torques from interaction of the rotor's patch--induced electric dipole moment with support voltages and the housing dipole moment. Far more important was the on--orbit discovery of forces and torques caused by interacting patches on the rotor and housing. Put simply, rotor and housing were spherical mechanically; electrically, they were not.

Three unforeseen effects emerged, differing in detail for each gyroscope: a changing polhode path and two patch effect torques. The changing polhode, originating in a $<1\,$pW dissipation of the rotor's kinetic energy, complicated the $C_g$ determination. Beginning with their body axes arbitrarily oriented and a 2 - 6 hr period, $T_p(t)$, the rotors transitioned to a final 1 - 4 hr period, each spinning nearly about its maximum inertia axis. The two  torques were 1) a spin-to-roll misalignment torque $200 - 500\times$ larger than predicted from mechanical asphericity, and 2) a `roll-polhode resonance' torque, where the roll averaging mentioned above would temporarily fail, making a particular gyroscope axis realign, or step over, by as much as 100 mas in 1--2 days when a high harmonic of its polhode rate came into resonance with spacecraft roll. The patch effect misalignment torque was discovered and quantified during the calibration phase by commanding the spacecraft to point to a series of positions at known angles to the guide star. 

Remarkably, a key to modeling all three  effects was {\it magnetic} asphericity: two patterns, magnetic fluxons and voltage patches, remained locked together in the rotor. A process of Trapped Flux Mapping (TFM) allowed exact tracking of the evolving polhode phase and angle needed for computing both $C_g(t)$ and torques. 

Further extensive and diversified post-science calibrations showed no significant disturbing effects other than the three just discussed.

\section{Data Analysis}

Two data analysis methods were used to determine and cross--check the relativity results. One, called `algebraic', was based on a parameter estimator utilizing the gyro dynamics and measurement models detailed below. It produced the primary science results (see Tables~\ref{tabRes},~\ref{tabError}, and Fig.~\ref{figEllipses}). The other method, called `geometric', though currently less precise, neatly eliminated the need to model the misalignment torque. This torque, being Newtonian, causes drift rates perpendicular to the misalignment vector $\vec\mu$. The drift in that direction is thus a combination of relativistic and Newtonian contributions, while the drift rate parallel to $\vec\mu$ -- when there is no roll--polhode resonance -- is pure relativity. The annual variation in misalignment direction allows determination of both relativity effects.
\vskip1mm
\noindent{\bf Gyro Dynamics}. The  $r_\mathrm{NS}$ (geodetic) and $r_\mathrm{WE}$ (frame--dragging) drift rates and two patch effect torques~\cite{keiser2009ssr} connect to the unit vector $\hat{s}(t)$ along the gyro spin axis by the equations of motion:
\bea
  \label{eqnS}
    \dot{s}_\mathrm{NS} = r_\mathrm{NS} + k(t) \mu_\mathrm{WE} + 
    \sum_m \left(a_m \cos \Delta \phi_m - b_m \sin \Delta \phi_m\right) \nonumber \\
  \dot{s}_\mathrm{WE}= r_\mathrm{WE} - k(t) \mu_\mathrm{NS}+
     \sum_m \left(a_m \sin \Delta \phi_m + b_m \cos \Delta \phi_m\right)  
\eea
The $k(t)\mu_\mathrm{WE},\;k(t)\mu_\mathrm{NS}$ terms give the misalignment drift rates,  $\mu_\mathrm{WE},\;\mu_\mathrm{NS}$ being the components of the misalignment vector $\vec\mu=\hat{\tau}-\hat{s}$, and $k(t)$ the polhode-dependent torque coefficient; $\hat{\tau}$ is the unit vector along the spacecraft roll axis. The resonance torques are governed by the third term of Eq.~(\ref{eqnS}) where $\Delta \phi_m(t)=\phi_r(t) - m \phi_p(t)$,  $\phi_r$ and $\phi_p$ being the known roll and polhode phases, with exact resonance occurring when the corresponding frequencies coincide, $\omega_r=m \omega_p$. To clarify the picture of a resonance, we keep only the resonance term on the right of Eq.~(\ref{eqnS}). Then the `step-over' in the NS---WE plane follows, to lowest order, a Cornu spiral winding out from its initial direction a day or so before the resonance, moving across, then winding back in to the new direction, its angular magnitude proportional to $\sqrt{a_m^2+b_m^2}$~\cite{keiser2009ssr}.
\vskip1mm
\noindent{\bf Science Data}. Determining $r_\mathrm{NS}$ and $r_\mathrm{WE}$ requires:\break 1) data from gyroscope, telescope, and roll reference signals (preprocessed and synchronized at $2\,$s intervals), with supporting TFM analyses based on separate $2200\,$~sample/s gyro readout signals; 2) the measurement model of Eq.~(\ref{mm}) below. The data came from 11 months of science operations, where spacecraft anomalies from events such as solar-flare-induced CPU reboots resulted in ten distinct but connectable data segments. Minor interruptions came from calibrations, transient electronics noise, and passages through the South Atlantic Anomaly. 
\vskip1mm
\noindent{\bf Measurement Model}.
With $\hat{s},\;\hat{\tau},\;\phi_r$ and $C_g(t)$ meanings already defined, we write SQUID signal $z(t)$ as:
\bea
z(t) = C_g(t) \,\left[(\tau_\mathrm{NS}-s_\mathrm{NS})\,
    \cos(\phi_r+\delta\phi)+\right. \nonumber\\
  \left.(\tau_\mathrm{WE}-s_\mathrm{WE})\,\sin(\phi_r+\delta\phi)\right]
    + \mathrm{bias} + \mathrm{noise};
\label{mm}
\eea
$(\tau_\mathrm{NS}-s_\mathrm{NS})$ and $(\tau_\mathrm{EW}-s_\mathrm{EW})$ are the NS and EW projections of the misalignment $\vec\mu=\hat{\tau}-\hat{s}$, and $\delta\phi$ is a constant roll phase offset.
\vskip1mm
\noindent{\bf Trapped Flux Mapping}. Central to the three main modeling challenges (determining $C_g(t)$ and handling  the misalignment and resonance
torques) was the information gained from Trapped Flux Mapping. The magnetic flux trapped in each rotor on cooling through its transition temperature formed a unique, stable fluxon pattern. By fitting to the spin harmonics of the $2200\,$sample/s data stream, we constructed detailed maps from which the three-dimensional orientation of each gyroscope could be tracked continuously over billions of turns. Rotor spin speed and spin-down rate were determined to $10\,$nHz and $1\,$pHz/sec, respectively~\cite{silbergleit2009ssr}. Crucial for data analysis were: 1) polhode phase $\phi_p(t)$  and angle $\gamma_p(t)$ both good to $1^\circ$ over the 353 days of science; and 2) detailed knowledge of the trapped flux contribution to $C_g(t)$.
\vskip1mm
\noindent{\bf TFM and $\bf{C_g(t)}$}. The scale factor $C_g(t)$ to be calibrated against aberration comes from the London moment $M_L$ aligned with the rotor's spin axis, plus a complex pattern of trapped flux $\sim10^{-2}M_L$ fixed in its body,
i.e. $C_g^{LM}+C_g^{TF}(t)$. To meet the necessary $<10^{-4}$ accuracy for $C_g(t)$ required connecting data from $\sim 5200$ successive guide star valid half--orbits enabled by TFM. 
\vskip1mm
\noindent{\bf TFM and Resonance Torques}. The role of TFM in computing the resonance torque was even more remarkable. Solving Eq.~(\ref{eqnS}) required precise knowledge of  $\phi_p(t)$  and $\gamma_p(t)$. It is curious to reflect that if GP-B had had an ideal London moment with no trapped flux, the computation would have been impossible.
\vskip1mm
\noindent{\bf Spacecraft Pointing}. To model the misalignment torque, the accurate value of $\hat{\tau}(t)$ is needed continuosly. During the half--orbit periods when the guide star is visible it is the sum of annual and orbital aberrations and minor terms including parallex, bending of stra light by the Sun, and pointing error from the telescope signal. For the occulted periods, $\hat{\tau}(t)$  was determined from the four science gyros' SQUID signals. The connection process required fast computation of the gyro spin vector $\hat{s}(t)$ through the entire 11 months of science data.
\vskip1mm
\noindent{\bf Parameter Estimator}. The science data set was processed by a nonlinear batch weighted least-squares (WLS) estimator implemented in iterative sequential information filter form~\cite{bier2006}. The cumulative information matrix for the full mission ($\sim1.4\times10^{7}$ data points per gyroscope) was calculated sequentially based on one-orbit data batches. To ensure robust convergence of numerous runs with different numbers of estimated parameters, we used the sigma-point filter technique~\cite{merw2004}. The lack of convergence observed in some runs of a standard WLS estimator was thus eliminated. Computations were critically facilitated by the replacement of differential equations~(\ref{eqnS}) with their analytical solution $s_\mathrm{NS}(t),\;s_\mathrm{WE}(t)$ for an arbitrary $k(t)$.
\begin{table}[b]
  \caption{\label{tabRes} Results}
  \begin{ruledtabular}
    \begin{tabular}{l c c}
      Source & $r_\mathrm{NS}$ (mas/yr) & $r_\mathrm{WE}$ (mas/yr) \\
      \hline \\[-2ex]
      Gyroscope 1 & $-6,588.6 \pm 31.7$ & $-41.3 \pm 24.6$ \\
      Gyroscope 2 & $-6,707.0 \pm 64.1$ & $-16.1 \pm 29.7$ \\
      Gyroscope 3 & $-6,610.5 \pm 43.2$ & $-25.0 \pm 12.1$ \\
      Gyroscope 4 & $-6,588.7 \pm 33.2$ & $-49.3 \pm 11.4$ \\
      \textbf{Joint (see text)} & $\mathbf{-6,601.8 \pm 18.3}$ 
        & $\mathbf{-37.2 \pm 7.2}$ \\[0.5ex]
      \hline \\[-2ex]
      GR prediction & $-6,606.1$ & $-39.2$ \\
      
    \end{tabular}
  \end{ruledtabular}
\end{table}

\section{Results and Conclusions}

\begin{figure}[t]
  \includegraphics[width = 9 cm]{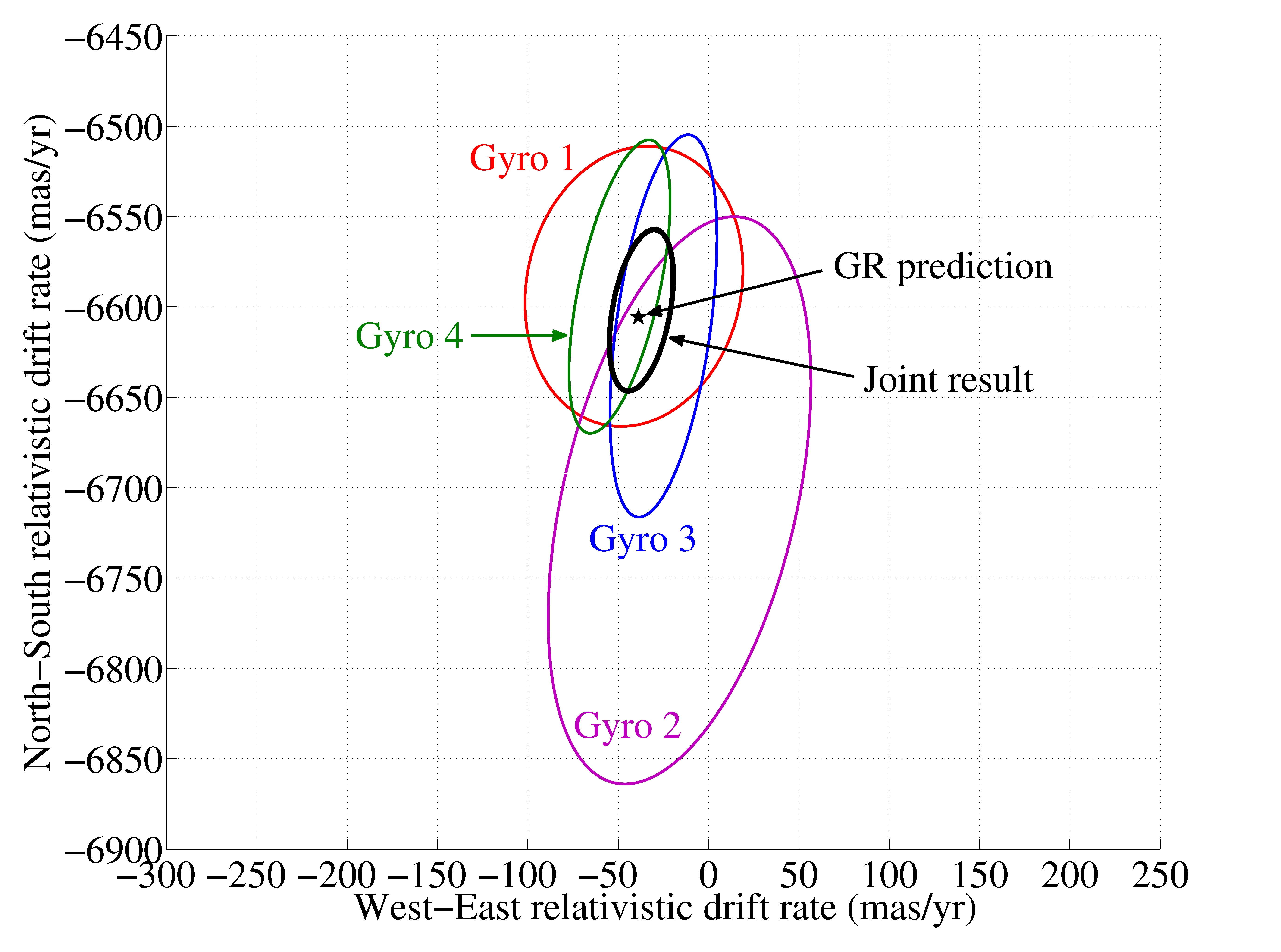}
  \caption{\label{figEllipses} North-South and West-East relativistic
    drift rate estimates (95\% confidence)
    for the four individual gyroscopes (colored ellipses)
    and the joint result (black ellipse).}
\end{figure}

The four gyroscope signals were analyzed independently and the results combined and cross-checked in various ways. Table~\ref{tabRes} lists
the four relativity estimates and final joint result with $1\,\sigma$ errors, also plotted as 95\% confidence
ellipses in Fig.~\ref{figEllipses}. The values  are referenced to inertial space by correcting for guide star proper motion, 
$27.3\pm0.1\,$mas/yr NS and $-20.0\pm0.1\,$mas/yr WE ~\cite{shapiro2011apj}, and the solar geodetic contribution of 
$7.3\pm0.3\,$mas/yr NS and $-16.2\pm0.6\,$mas/yr WE. 

The result for each gyroscope is the average of 10 drift rate estimates
based on 10 distinct parameter sets determined by the following uniform procedure.
The submodels for scale factor $C_g(t)$, misalignment torque coefficient $k(t)$, polhode phase $\phi_p(t)$, etc., are linear combinations of certain basis functions with coefficients to be estimated.
The number of terms in each submodel is increased from zero
until the change in the relativistic drift rate estimates becomes less than $0.5\,\sigma$. 
This defines the baseline set for a given gyroscope. Table~\ref{tabRes} gives the weighted mean of the
baseline and nine `sensitivity run' estimates for each gyroscope,
whereby in each run the number of terms in a single submodel is increased by 1
above the baseline value. The post-fit residuals were white with $\chi^2 \approx 1$ for
all gyroscopes. 

The joint 4--gyro result is a weighted average of the four individual drift rates using their $2\times2\; r_\mathrm{NS},\,r_\mathrm{WE}$ sub-blocks of the full covariance matrix for each gyroscope. In the NS direction the scatter of the individual estimates is 28\% larger than the individual uncertainties indicate, while in the WE direction it is 37\% smaller. The individual and combined statistical uncertainties are corrected for their `over'  and `under' dispersion using the $\chi^2$ of the individual estimates in the NS and WE directions. The `parameter sensitivity' covariance matrix is the scatter of all $10^4$ combinations of drift rate estimates obtained in $10$ sensitivity runs for each of the 4 gyroscopes. The total covariance matrix is a sum of three matrices: the corrected statistical one, the sensitivity one, and the covariance matrix of unmodeled systematic effects.
The $1\,\sigma$ uncertainties of the joint result shown in Table II are the square roots of the diagonal elements of the total covariance matrix.

Table~\ref{tabError} summarizes uncertainties of the joint 4-gyro result.
The three main disturbances (evolving scale factor and the two Newtonian torques) 
are modeled and thus contribute to the statistical uncertainty. The systematic uncertainty in Table~\ref{tabError} includes: 1) uncertainty from the sensitivity analysis to the number of model parameters; and 2) small effects not incorporated in the model. 

We performed several important data analysis cross-checks. First, the gyro drift rate results of Table~\ref{tabRes} are confirmed by separate analyses of the segmented data. The 24 independent results from six data segments for each gyroscope are all consistent with the joint result within their confidence limits, demonstrating the internal consistency of the model. Second, the misalignment torque coefficients $k$ determined during the calibration phase proved to be in excellent agreement with the end-of-mission values estimated by both algebraic and geometric analysis methods. No less impressive was the agreement between the time history $k(t)$ throughout the mission obtained by the two methods. As for the roll-polhode resonance torques, the gyro dynamics model of Eq.~(\ref{eqnS}) predicts that during a resonance the gyroscope orientation axis approximately follows a Cornu spiral. Indeed, that is typically observed in orbit-by-orbit gyro orientations determined by both data analysis methods. 

Further cross--checks from the geometric method included relativity estimates for 2 of the 4 gyroscopes, in both NS and WE directions, in statistical agreement with the algebraic results, and an estimate of the gravitational deflection of light by the Sun. IM Pegasi's closest approach to the Sun occurs on March 11th, with ecliptic latitude $22.1^\circ$ as compared with the grazing incidence value $0.265^\circ$ in classical light deflection tests. The maximum deflection predicted by GR is $21.7\,$mas; the observed $21\pm7\,$mas serves as a neat structural confirmation of Gravity Probe B.

Lunar laser ranging has reported a measurement of the de~Sitter solar geodetic effect to 0.7\%~\cite{williams1996}.
Analyses of laser ranging to the LAGEOS and LAGEOS II spacecraft report a 10\% - 30\% measurement of the  frame-dragging effect, assuming the GR value for the geodetic precession~\cite{ciufolini2004nat, iorio2009ssr}.  GP-B provides independent measurements of the geodetic and frame-dragging
effects at an accuracy of 0.28\% and 19\%, respectively.
\begin{table}[t]
  \caption{\label{tabError} Contributions to Experiment Uncertainty}
  \begin{ruledtabular}
    \begin{tabular}{l c c}
      Contribution & NS (mas/yr) & WE (mas/yr) \\
      \hline
      \\[-2ex]
      \textbf{Statistical (modeled)} & $\mathbf{16.8}$ & $\mathbf{5.9}$ \\[1ex]
      \textbf{Systematic (unmodeled)} & & \\
      Parameter sensitivity & 7.1 & 4.0 \\[1ex]
      Guide star motion & 0.1 & 0.1 \\
      Solar geodetic effect & 0.3 & 0.6 \\
      Telescope readout & $0.5$ & $0.5$ \\
      Other readout uncertainties & $<1$ & $<1$ \\
      Other classical torques & $<0.3$ & $<0.4$ \\
            \textbf{Total Stat. + Sys.} & $\mathbf{18.3}$ & $\mathbf{7.2}$ \\
    \end{tabular}
  \end{ruledtabular}
\end{table}

\section{Acknowledgements}

This work was supported by NASA Contract NAS8-39225 through Marshall Space Flight Center until Jan. 2008 and by King Abdulaziz City for Science and Technology (KACST) since Sept. 2008. The Jan.---Sept. 2008 funding was through a personal gift by Richard Fairbank with matching funds from NASA and Stanford University. A personal gift from Vance and Arlene Coffman with matching funds by KACST was crucial to finalizing 
this paper. Support from ICRA/ICRANet is also gratefully acknowledged. GP-B would never have succeeded without the work of hundreds of people at Stanford, NASA, Lockheed--Martin, and  elsewhere. Some 500 graduate, undergraduate and high school students have contributed much to the project. Our special thanks go to members of the NASA-appointed Science Advisory Committee, chaired by Clifford Will, for their support and advice, and to Charbel Farhat for all his help.

\bibliography{GPB_PRL2010bib}

\begin{thebibliography}{17}
\expandafter\ifx\csname natexlab\endcsname\relax\def\natexlab#1{#1}\fi
\expandafter\ifx\csname bibnamefont\endcsname\relax
  \def\bibnamefont#1{#1}\fi
\expandafter\ifx\csname bibfnamefont\endcsname\relax
  \def\bibfnamefont#1{#1}\fi
\expandafter\ifx\csname citenamefont\endcsname\relax
  \def\citenamefont#1{#1}\fi
\expandafter\ifx\csname url\endcsname\relax
  \def\url#1{\texttt{#1}}\fi
\expandafter\ifx\csname urlprefix\endcsname\relax\def\urlprefix{URL }\fi
\providecommand{\bibinfo}[2]{#2}
\providecommand{\eprint}[2][]{\url{#2}}

\bibitem[{\citenamefont{Schiff}(1960)}]{schiff1960}
\bibinfo{author}{\bibfnamefont{L.~I.} \bibnamefont{Schiff}},
  \bibinfo{journal}{Phys. Rev. Lett.} \textbf{\bibinfo{volume}{4}},
  \bibinfo{pages}{215} (\bibinfo{year}{1960}).

\bibitem[{\citenamefont{{de Sitter}}(1916)}]{deSitter1916}
\bibinfo{author}{\bibfnamefont{W.}~\bibnamefont{{de Sitter}}},
  \bibinfo{journal}{MNRAS} \textbf{\bibinfo{volume}{77}}, \bibinfo{pages}{155}
  (\bibinfo{year}{1916}).

\bibitem[{\citenamefont{{Lense} and {Thirring}}(1918)}]{lense1918pz}
\bibinfo{author}{\bibfnamefont{J.}~\bibnamefont{{Lense}}} \bibnamefont{and}
  \bibinfo{author}{\bibfnamefont{H.}~\bibnamefont{{Thirring}}},
  \bibinfo{journal}{Phys. Zeits.} \textbf{\bibinfo{volume}{19}},
  \bibinfo{pages}{156} (\bibinfo{year}{1918}).

\bibitem[{\citenamefont{{Thorne}}(1988)}]{thorne1988}
\bibinfo{author}{\bibfnamefont{K.~S.} \bibnamefont{{Thorne}}}, in
  \emph{\bibinfo{booktitle}{Near Zero: New Frontiers of Physics}}, edited by
  \bibinfo{editor}{\bibfnamefont{J.~D.} \bibnamefont{{Fairbank}}}
  \bibnamefont{et~al.} (\bibinfo{publisher}{W.H.Freeman and Co.},
  \bibinfo{address}{New York}, \bibinfo{year}{1988}), pp.
  \bibinfo{pages}{573--586}.

\bibitem[{foo()}]{footnote1}
\bibinfo{note}{The inertial axes used in the paper: West--East (WE) - along the
  cross-product of the unit vector to the guide star and the unit vector $\hat
  z$ of the inertial JE2000 frame; North--South (NS) - along the cross-product
  of the WE unit vector and the unit vector to the guide star.}

\bibitem[{\citenamefont{{Pugh}}(1959)}]{pugh1959}
\bibinfo{author}{\bibfnamefont{G.~E.} \bibnamefont{{Pugh}}},
  \bibinfo{type}{Tech. Rep.} (\bibinfo{year}{1959}), \bibinfo{note}{{reprinted
  in: Ruffini, R.J., Sigismondi, C. (Eds.), {\it Nonlinear Gravitodynamics. The
  Lense-Thirring Effect. World Scientific}, Singapore, pp. 414-426}}.

\bibitem[{\citenamefont{{DeBra}}(1988)}]{debra1988}
\bibinfo{author}{\bibfnamefont{D.~B.} \bibnamefont{{DeBra}}}, in
  \emph{\bibinfo{booktitle}{Near Zero: New Frontiers of Physics}}, edited by
  \bibinfo{editor}{\bibfnamefont{J.~D.} \bibnamefont{{Fairbank}}}
  \bibnamefont{et~al.} (\bibinfo{publisher}{W.H.Freeman and Co.},
  \bibinfo{address}{New York}, \bibinfo{year}{1988}), pp.
  \bibinfo{pages}{691--699}.

\bibitem[{\citenamefont{{Darling}}(1989)}]{darling1989}
\bibinfo{author}{\bibfnamefont{T.~W.} \bibnamefont{{Darling}}},
  \bibinfo{journal}{School of Physics} p.~\bibinfo{pages}{88}
  (\bibinfo{year}{1989}).

\bibitem[{\citenamefont{{Speake}}(1996)}]{speake1996cqg}
\bibinfo{author}{\bibfnamefont{C.~C.} \bibnamefont{{Speake}}},
  \bibinfo{journal}{Classical and Quantum Gravity}
  \textbf{\bibinfo{volume}{13}}, \bibinfo{pages}{A291} (\bibinfo{year}{1996}).

\bibitem[{\citenamefont{{Keiser} et~al.}(2009)\citenamefont{{Keiser},
  {Kolodziejczak}, and {Silbergleit}}}]{keiser2009ssr}
\bibinfo{author}{\bibfnamefont{G.~M.} \bibnamefont{{Keiser}}},
  \bibinfo{author}{\bibfnamefont{J.}~\bibnamefont{{Kolodziejczak}}},
  \bibnamefont{and} \bibinfo{author}{\bibfnamefont{A.~S.}
  \bibnamefont{{Silbergleit}}}, \bibinfo{journal}{Space Science Reviews}
  \textbf{\bibinfo{volume}{148}}, \bibinfo{pages}{383} (\bibinfo{year}{2009}).

\bibitem[{\citenamefont{{Silbergleit} et~al.}(2009)\citenamefont{{Silbergleit},
  {Conklin}, {DeBra}, and {others}}}]{silbergleit2009ssr}
\bibinfo{author}{\bibfnamefont{A.}~\bibnamefont{{Silbergleit}}},
  \bibinfo{author}{\bibfnamefont{J.}~\bibnamefont{{Conklin}}},
  \bibinfo{author}{\bibfnamefont{D.}~\bibnamefont{{DeBra}}}, \bibnamefont{and}
  \bibinfo{author}{\bibnamefont{{others}}}, \bibinfo{journal}{Space Science
  Reviews} \textbf{\bibinfo{volume}{148}}, \bibinfo{pages}{397}
  (\bibinfo{year}{2009}).

\bibitem[{\citenamefont{{Bierman}}(2006)}]{bier2006}
\bibinfo{author}{\bibfnamefont{G.~J.} \bibnamefont{{Bierman}}},
  \emph{\bibinfo{title}{{Factorization Methods for Discrete Sequential
  Estimation}}} (\bibinfo{publisher}{Dover Publications, Inc.},
  \bibinfo{address}{New York}, \bibinfo{year}{2006}), \bibinfo{edition}{2nd}
  ed.

\bibitem[{\citenamefont{{van der Merwe} et~al.}(2004)\citenamefont{{van der
  Merwe}, {Wan}, and {Julier}}}]{merw2004}
\bibinfo{author}{\bibfnamefont{R.}~\bibnamefont{{van der Merwe}}},
  \bibinfo{author}{\bibfnamefont{E.}~\bibnamefont{{Wan}}}, \bibnamefont{and}
  \bibinfo{author}{\bibfnamefont{S.~J.} \bibnamefont{{Julier}}}, in
  \emph{\bibinfo{booktitle}{Proceedings of the AIAA Guidance, Navigation \&
  Control Conference, Providence, RI}} (\bibinfo{year}{2004}), pp.
  \bibinfo{pages}{16--19}.

\bibitem[{\citenamefont{{Shapiro} et~al.}(2011)\citenamefont{{Shapiro},
  {Bartel}, {Bietenholz}, and {others}}}]{shapiro2011apj}
\bibinfo{author}{\bibfnamefont{I.~I.} \bibnamefont{{Shapiro}}},
  \bibinfo{author}{\bibfnamefont{N.}~\bibnamefont{{Bartel}}},
  \bibinfo{author}{\bibfnamefont{M.~F.} \bibnamefont{{Bietenholz}}},
  \bibnamefont{and} \bibinfo{author}{\bibnamefont{{others}}},
  \bibinfo{journal}{submitted to Ap.J.}  (\bibinfo{year}{2011}).

\bibitem[{\citenamefont{{Williams} et~al.}(1996)\citenamefont{{Williams},
  {Newhall}, and {Dickey}}}]{williams1996}
\bibinfo{author}{\bibfnamefont{J.~G.} \bibnamefont{{Williams}}},
  \bibinfo{author}{\bibfnamefont{X.~X.} \bibnamefont{{Newhall}}},
  \bibnamefont{and} \bibinfo{author}{\bibfnamefont{J.~O.}
  \bibnamefont{{Dickey}}}, \bibinfo{journal}{Phys. Rev. D}
  \textbf{\bibinfo{volume}{93}}, \bibinfo{pages}{6730} (\bibinfo{year}{1996}).

\bibitem[{\citenamefont{{Ciufolini} and {Pavlis}}(2004)}]{ciufolini2004nat}
\bibinfo{author}{\bibfnamefont{I.}~\bibnamefont{{Ciufolini}}} \bibnamefont{and}
  \bibinfo{author}{\bibfnamefont{E.~C.} \bibnamefont{{Pavlis}}},
  \bibinfo{journal}{Nature} \textbf{\bibinfo{volume}{431}},
  \bibinfo{pages}{958} (\bibinfo{year}{2004}).

\bibitem[{\citenamefont{{Iorio}}(2009)}]{iorio2009ssr}
\bibinfo{author}{\bibfnamefont{L.}~\bibnamefont{{Iorio}}},
  \bibinfo{journal}{Space Science Reviews} \textbf{\bibinfo{volume}{148}},
  \bibinfo{pages}{363} (\bibinfo{year}{2009}).

\end{thebibliography}

\end{document}